\documentclass[aps,prb,twocolumn]{revtex4}

\usepackage{epsfig}
\usepackage{graphicx}
\usepackage{amssymb}
\usepackage{mathrsfs}
\usepackage{dcolumn}

\newcommand{\pr}{\partial}

\newcommand{\p}{\prime}
\newcommand{\om}{\omega}
\newcommand{\ra}{\rangle}
\newcommand{\la}{\langle}

\begin{document}

\title{Non-linear Nyquist theorem: A conjecture}
\author{Navinder Singh}
\affiliation{Physical Research Laboratory, Ahmedabad, India, PIN: 380009.}

\begin{abstract}
Thermodynamics of equilibrium states is well established. However, in nonequilibrium few general
results are known.  One prime and important example is that of Nyquist theorem. It relates
equilibrium tiny voltage fluctuations across a conductor with its resistance. In linear systems it
was proved in its generality in a beautiful piece of work by Callen and Welton
(in 1950s\cite{cw}). However Callen-Welton's formalism has not been extended to nonlinear systems up to now, although alternative methods exist (like Kubo's approach) that leads to formal and implicit expressions {\it at nonlinear order} with no practical consequence.  Here--using a brute-force method--we conjecture  "a non-linear Nyquist theorem". This is an explicit formula much like Nyquist's original one.  Our conjecture  is based upon tests of the conjectured explicit formula in specific systems.  We conjecture that higher moments of {\it equilibrium} fluctuations bear a relation to {\it nonlinear} admittance very similar to Nyquist's relation. Thus one can easily compute nonlinear admittance from the character of {\it equilibrium} fluctuations. Our relation will have great practical applicability, for example for electronic devices that operate under nonlinear response.
\end{abstract}
\pacs{............}

\maketitle

By a brute-force method an extension of Callen and Welton's seminal work of 1951\cite{cw} is developed. Before we present that,  we review the seminal works of Nyquist\cite{nyq} and of Callen-Welton\cite{cw}. 
\begin{figure}[h!]
\includegraphics[height = 5cm, width = 6cm]{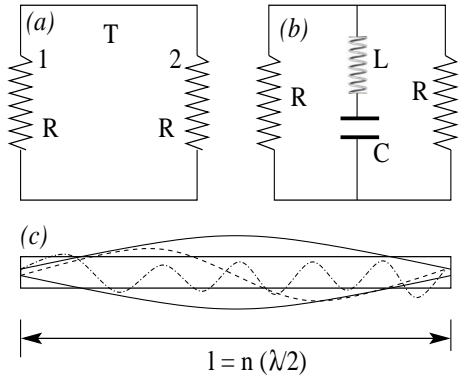}
\caption{(a) Two resistances connected with each other. Both are at same temperature $T$. (b) Arrangement to show that at any frequency equal power flows from both sides. (c) Standing modes in the shunted transmission line.}
\label{f1}
\end{figure}
If you connect a resistance with a sensitive voltmeter or oscilloscope you observe an ac voltage of
zero mean but non-zero Root Mean Square (RMS) value of very small magnitude (sub-micro volts in
ordinary conductors). This thermally induced  noise voltage (the Johnson-Nyquist noise) was first
observed by John B. Johnson of Bell Labs in 1926 and its theoretical explanation was provided by his
colleague Harry Nyquist in 1927. Hence called Johnson-Nyquist (JN) noise.  Nyquist proved that the
frequency integrated variance of  voltage is proportional to the frequency integrated resistance:
$\la V^2 \ra = 4 k_B T \int_0^\infty R(\om) d \om$. He proved this by using simple but ingenious
thermodynamical arguments\cite{nyq}. He first shows that if you connect two {\it equal} resistances
in parallel with each other (figure 1(a)), then the power dissipated in first resistance due to JN
noise voltage produced by the second resistance  will be equal to power dissipated in the second due
JN noise voltage produced by the first, provided both resistances are kept at the same temperature.
Thus there is a balance of power flow. This is in accordance with second law of thermodynamics. He
further refines it by showing that the above inference is true at all frequencies\cite{nyq} thus
obtaining an important conclusion that RMS voltage must be a universal function of frequency. This
he shows by connecting a $LC$ circuit in between the resistances (figure 1(b)). $LC$ circuit acts
like a shunt at a specific frequency $\om=\frac{1}{\sqrt{LC}}$. If the power transfer is non uniform
in frequency, then more power will be shunted from 1 to 2 as compared to that from 2 to 1, or vice
versa. Thus the arrangement in figure 1 (b) will spontaneously leads to heating of one resistance as
compared to the other, again violating the second law of thermodynamics. Thus, there is a balance of
power flow in each frequency interval and power transferred must be a universal function of
frequency. 

It was know from the experiments of Johnson that the JN noise is a universal function of resistance
and temperature (it does not depend on the material of the resistance i.e. whether it is carbon
resistance or metallic). Thus $V_{rms} =f(R,T,\om)$. Where $f$ is some universal function. Nyquist
explicitly derives this universal function using the law of equipartition of energy and counting the
number of standing modes in a transmission line. 

In his thought experiment he first disconnects the resistances and then connects them at the ends of a lossless transmission line for a time interval $l/v$. Where $l$ is the length of the transmission line and $v$ is speed of electromagnetic waves in the line. After this, he removes the resistances and short circuits the ends of the transmission line. This leads to standing modes in the transmission line (figure 1(c)). Number of standing modes in the frequency interval $f$ to $f+df$ will be $(2l/v) df$. By equipartition of energy each mode has $k_B T$ of energy. Thus energy in the frequency interval  $f$ to $f+df$ will be $k_B T (2l/v) df$. Average power transferred in time $l/v$ is $2 k_bT df$ (energy/time). For a linear system average power is $\frac{\la V^2\ra_f}{2 R} df$ (where $\la V^2\ra_f$ is variance per unit frequency). Thus leading to $\la V^2\ra_f df = 4 k_B T R df$. If the resistance is frequency dependent then frequency integrated variance bears the relation 
\begin{equation}
\la V^2 \ra = 4 k_B T \int_0^\infty R(\om) d \om,
\label{nyq}
\end{equation}
known as Nyquist theorem\cite{nyq}. {\it{The fundamental importance of this relation and other
general linear fluctuation-dissipation theorems\cite{fdts} is that the equilibrium fluctuations (for
example, JN noise) has "hidden" information regarding transport coefficients (here the resistance
$R$).}} This fact is the cornerstone of liner nonequilibrium statistical mechanics\cite{lnsm}. 

Callen and Welton gave rigorous quantum mechanical foundation to Nyquist's result. They consider a
conductor of length $L$ and of impedance $Z(\om)$ biased with an ac voltage $V(t) =V_0 \sin(\om
t)$\footnote{Here, the electrical circuit case is an illustrative example. Applicability of these
relations is much more general\cite{cw}}. The total Hamiltonian $H$ of the system composes two
parts: $H_0$ the unperturbed part, and $V(t) \hat{Q}$ the perturbation. That is $H =
H_0+V(t)\hat{Q}$. Where $\hat{Q} = \sum_i e \frac{\hat{x}_i}{L}$ and $x_i$ is the position of
$i^{th}$ particle from one end of the conductor ($\hat{x}_i$ is an operator corresponding to that).
They assume that eigensystem of $H_0$ is known with $E_n$ as eigen energies and $\phi_n$ as
eigenfunctions. Compute an average power absorbed by the conductor from the battery. For this, let
the wavefunction of the perturbed system is $\psi(t) =\sum_n a_n(t)\phi_n$ which obey the
Schroedinger's equation $i \hbar \frac{\pr \psi(t)}{\pr t} = (H_0+ V(t)\hat{Q}) \psi(t)$. Under the
assumption of weak perturbation (which is usually the case) the expansion coefficients $a_n(t)$ are
expressed as perturbation series. Retaining up to the first order, the transition probability to
find the system in some final state $|f\ra$ of energy $E_f$ at time $t$ when the system was in state
$|i\ra$ at an initial time $t=0$ is calculated to be $\mathscr{P}_{if}(t) = |\la f|\psi(t)\ra|^2 =
\frac{1}{2} \pi t \frac{V_0^2}{\hbar}|Q_{E_f,E_i}|^2 (\delta(\om+\om_{if})
+\delta(\om-\om_{if}))$\cite{cw}. Where the notation  $Q_{E_f,E_i}$ means matrix element $\la
E_f|\hat{Q}|E_i\ra$.

If the final states form a continuum (generally true for a system in thermodynamic limit) then the total transition probability from initial state $|i\ra$ to {\it any final state} per unit time is  $\mathscr{P}_i = \frac{1}{t}\sum_f \mathscr{P}_{if} = \frac{1}{t}\int dE_f \rho(E_f) \mathscr{P}_{if} = \frac{\pi}{2}\frac{V_0^2}{\hbar}(|\la E_i+\hbar \om|\hat{Q}|E_i\ra|^2\rho(E_i+\hbar \om) + |\la E_i-\hbar \om|\hat{Q}|E_i\ra|^2\rho(E_i-\hbar \om))$. While replacing sums by integrals we introduce density of states $\rho(E)$ (number of states per unit energy). Two terms in the parenthesis has the following physical meaning. First term (with $\la E_i+\hbar\om |\hat{Q}| E_i\ra$) represents photon absorbed by the system (conductor) from the battery in which initial state with energy $E_i$ changes to a state with energy $E_i+\hbar\om$. The other term represents loss of a photon by the system (i.e., $\la E_i-\hbar \om|\hat{Q}|E_i\ra$). Thus power absorbed is $Power(E_i)=\hbar \om  \times \frac{\pi}{2}\frac{V_0^2}{\hbar}(|\la E_i+\hbar \om|\hat{Q}|E_i\ra|^2\rho(E_i+\hbar \om) - |\la E_i-\hbar \om|\hat{Q}|E_i\ra|^2\rho(E_i-\hbar \om))$. As $\hbar \om$ is the energy of a single photon. Also notice the minus sign between the two terms in the parenthesis that represents gain minus loss ($gain-loss$). The {\it average} power absorbed at frequency $\om$ is $\la Power\ra_\om = \int_0^\infty dE_i \rho(E_i) f(E_i) Power(E_i)$ and is obtained by summing over all possible initial states with thermodynamic weighting factor $f(E_i) = e^{-E_i/k_B T}$. For a linear system $\la Power\ra_\om = \frac{1}{2} V_0^2 \frac{R(\om)}{|Z(\om)|^2} =\frac{1}{2} V_0^2 Y(\om) $,  and from this they obtain an expression for  $Y(\om)$ (the real part of linear admittance).

Next, they analyze the nature of equilibrium voltage fluctuations in the conductor (when battery is disconnected). Average current $\dot{\la\hat{Q}}\ra = \frac{i}{\hbar} \la E_n |[H_o,\hat{Q}]|E_n\ra$ vanishes in equilibrium (as expected), but its square does not. With a simple calculation they show that $\la|\dot{\hat{Q}}^2|\ra = \int_0^\infty d\om \hbar \om^2 \int_0^\infty dE \rho(E) f(E) (|\la E_i+\hbar \om|\hat{Q}|E_i\ra|^2\rho(E_i+\hbar \om) + |\la E_i-\hbar \om|\hat{Q}|E_i\ra|^2\rho(E_i-\hbar \om))$. With the expression $V(\om) = |Z(\om)| \dot{\hat{Q}}(\om)$ variance of the voltage $\la V^2\ra$ can be expressed in terms of the matrix elements and Density Of States (DOS).  From these expressions of admittance and variance of equilibrium voltage it can be shown that $\la V^2\ra = \frac{2}{\pi}\int_0^\infty d\om R(\om) E(\om,T)$ with $E(\om,T) = \hbar \om/2+\hbar \om/(e^{\hbar \om/k_B T} - 1)$. This is known as the Callen-Welton's theorem\cite{cw}. In the high temperature limit $k_B T\gg\hbar \om,~~E(\om,T)$ can approximated by $k_B T$. This leads to  $\la V^2\ra = \frac{2}{\pi} k_B T \int_0^\infty d\om R(\om)$ which is nothing but the Nyquist's result now with correct coefficient! 

A survey of literature: Callen-Welton's theorem and its statistical mechanical formulation by
Kubo\cite{fdts} completed the program of linear Fluctuation-Dissipation Theorems (FDT) by 1957 (
i.e., connecting transport coefficients of linear irreversible processes (for example, Ohm's law:
current proportional to voltage) with {\it equilibrium} fluctuations). The next logical step was to
extend these theorems of Nyquist, Callen-Welton, and Kubo to nonlinear regime (in which, for
example, current is also proportional to higher powers of voltage). First steps in these directions
were taken by William Bernard and Herbert Callen\cite{rev1}, and by Russian investigators: R. L.
Stratonovich\cite{stra, strabook}; G. F. Efremov\cite{efre};  G. N. Bochkov and Yu. E.
Kuzovlev\cite{bk}. 

In Bernard-Callen's work\cite{rev1} an expression for nonlinear FDT is given (equation (162) in\cite{rev1}). However, as they point out, it does not constitute a thermodynamical relation as their function $f_{jki}^{(0)}(\om_1,\om_2)$ is not macroscopically observable (see discussion below equation (162) in\cite{rev1}). In general at the nonlinear order admittance and fluctuation expressions becomes extremely complicated and this algebraic complexity hindered the progress. But Russian investigators were able to make progress by exploiting the principle of time reversal invariance\cite{strabook}.  In 1967 Stratonovich derives nonlinear FDT under the Markovian assumption\cite{strabook} and using a master equation. In 1968, going beyond the Markovian limit, Efremov proves the non-Markovian quadratic FDT\cite{strabook}. The expressions (called three-subscript and four-subscript relations) obtained by these investigators are highly formal and implicit (see, for example, equation (6.1.88) in\cite{strabook} for fourfold correlator). In 1977, Bochkov and Kuzovlev, again by exploiting the principle of time reversal invariance of microscopic dynamics, develop a general theory of thermal fluctuations in nonlinear systems\cite{bk}. They obtain a formula (equation (4) in\cite{bk}) that characterizes the excitation of the system from the state of thermodynamic equilibrium. From this fundamental formula\footnote{This formula is the genesis of the recent "Fluctuation Theorems\cite{bkrev}".} they obtain three and four index relations between the equilibrium and nonequilibrium moment functions. Again, these formulae suffer from analytical complexity and a direct and explicit analogy with Nyquist theorem is difficult to obtain. 

In the present investigation we obtain a direct generalization the the linear Nyquist's theorem
without using any master equation.  Our method is a brute-force extension of the original
Callen-Welton result. End result is an explicit and compact formula much like Nyquist's original
one. In addition, with the present approach, we obtain an explicit expression for nonlinear
admittance in terms of density-of-states of a system and current matrix elements. Thus, present work
will also be useful {\it in direct calculations} of nonlinear admittances.

With this physical background and a survey of literature, we now motivate a nonlinear Nyquist
theorem. We start with the setting  used by Callen and Welton (a conductor biased with a battery).
The total Hamiltonian is $H = H_0+V(t)\hat{Q}$ where $\hat{Q} = \sum_i e \frac{\hat{x}_i}{L}$ as
before. Our aim is to compute the average power absorbed by our  system (the conductor) at the next
order of the applied voltage. To the second order in perturbation theory the expansion coefficient
$a_n(t)$ of $\psi(t) =\sum_n a_n(t)\phi_n$  can be written as
\begin{widetext}
\begin{equation}
b_n(t) = (-i/\hbar)^2\sum_m \int_0^t dt^\p H^\p_{nm}(t^\p) \int_0^{t^\p} dt^{\p\p} H^\p_{mi}(t^{\p\p}) e^{-i(\om_{mn}t^\p+\om_{in}t^{\p\p})} .
\end{equation}
Here $H^\p_{nm}(t) = V_0 \sin(\om t) \hat{Q}$ and $e^{-i E_n t/\hbar} b_{n}(t)$. This leads to the
transition probability ($\mathscr{P}_{fi} =|\la f|\psi(t)\ra|^2$):
\begin{eqnarray}
\mathscr{P}_{fi} &=& (V_0/\hbar)^4\sum_{m,n} Q_{fn} Q_{ni} Q^\ast_{fm} Q^\ast_{mi} \int_0^t dt^\p \sin(\om t^\p) e^{-i \om_{nf} t^\p} \int_0^{t^\p} dt^{\p\p} \sin(\om t^{\p\p}) e^{- i \om_{in} t^{\p\p}} \nonumber\\
&\times&
\int_0^t dt^\p \sin(\om t^\p) e^{-i \om_{mf} t^\p} \int_0^{t^\p} dt^{\p\p} \sin(\om t^{\p\p}) e^{- i \om_{im} t^{\p\p}}.
\label{3}
\end{eqnarray}
With tedious algebra (see supplementary information), this can be simplified to
\begin{eqnarray}
\mathscr{P}_{fi} &=& (\pi t/8)(V_0/\hbar)^4\sum_{m,n} Q_{fn} Q_{ni} Q^\ast_{fm} Q^\ast_{mi} \left(\frac{\delta(2\om-\om_{if})}{(\om-\om_{in})(\om-\om_{im})}+\frac{\delta(2\om+\om_{if})}{(\om+\om_{in})(\om+\om_{im})}\right) \nonumber\\
&+&
(\pi t/2)(V_0/\hbar)^4\sum_n |Q_{fn}|^2 |Q_{ni}|^2 \left(\frac{\om}{\om^2-\om_{in}^2}\right)^2 (\delta(\om-\om_{nf})+\delta(\om+\om_{nf})).
\label{4}
\end{eqnarray} 
\end{widetext}
By converting sums into integrals $\sum_n \longrightarrow \int dE \rho(E)$ it is possible to do
integrals over the final states using the properties of the Dirac delta functions and then by
recognizing the emission and absorption process the average power transferred (from battery to
conductor) can expressed as
\begin{widetext}
\begin{eqnarray}
\la Power_\om\ra^{(4)} &=& (\pi V_0^4\om/4)\int_0^\infty dE_i \rho(E_i) f(E_i) \int_0^\infty dE_m \rho(E_m) \int_0^\infty dE_n \rho(E_n)\nonumber\\
&\times& \left(\frac{\rho(E_i+2\hbar \om) Q_{E_i+2\hbar \om,E_n} Q_{E_n,E_i} Q^\ast_{E_i+ 2 \hbar \om,E_m} Q^\ast_{E_m,E_i} }{(\hbar \om +E_i -E_n)(\hbar\om +E_i-E_m)} \right. \nonumber \\ 
&-& \left. \frac{\rho(E_i-2\hbar \om) Q_{E_i-2\hbar \om,E_n} Q_{E_n,E_i} Q^\ast_{E_i-2 \hbar \om,E_m} Q^\ast_{E_m,E_i} }{(\hbar \om -E_i +E_n)(\hbar\om -E_i+E_m)}  \right).
\label{5}
\end{eqnarray}
  \end{widetext}
Technical details are given in supplementary information. For nonlinear system under consideration $\la Power_\om\ra^{(4)} = \frac{3}{8}V_0^4 Y_R^{(4)}(\om)$ where $Y_R^{(4)}(\om)$ is the real part of the nonlinear admittance. Thus real part of the nonlinear admittance can be expressed in terms of matrix elements of $Q$ and Density Of States (DOS). {\it An important difference at this nonlinear order is that the system absorbs two quanta ($2\hbar\om$) from battery while in the linear order it absorbs one quantum ($\hbar\om$)}. Notice the matrix elements $Q_{E_i+2\hbar\om,E_n} Q_{E_n,E_i}$.

When battery is disconnected from the conductor the conductor regains equilibrium in some relaxation time. In equilibrium, voltage has tiny fluctuations, mean of odd powers of voltage vanishes but mean of even powers  does not. In the present case we need to compute an average of the fourth power of the current operator $\la \dot{\hat{Q}}^4\ra = \sum_n f(E_n) \la E_n |\dot{\hat{Q}}^4| E_n\ra$. The expression in terms of the DOS and matrix elements is extremely lengthy and is given in the supplementary information (equation (19)). Due to horrendous analytical complexity, it is not possible to establish a direct relation between $Y_R^{(4)}(\om)$ and $\la \dot{\hat{Q}}^4\ra$ as was done by Callen and Welton in the linear regime where the formulae  were simpler.

We adopt the following  strategy. We simplify the above expressions for $Y_R^{(4)}(\om)$ and $\la \dot{\hat{Q}}^4\ra$ by considering two simple physical systems and test whether there exists a relation analogous to the Nyquist relation or not. In the first system we consider constant DOS and constant matrix elements and in the second system we consider constant DOS but (non)constant matrix elements which are obtained by an explicit calculation. For the first system let us assume that $\rho(E) = \rho_0 \Theta(E_{up}-E)$ (step function DOS) and $Q_{E_i, E_j} = \eta e$. Here $E_{up}$ is the upper cut-off in the DOS, and $\eta$ is dimensionless constant and $e$ is the electronic charge. Nonlinear admittance (equation (17), supplementary information) in this case can be simplified to
 \begin{widetext}
\begin{eqnarray}
Y_R^{(4)}(\om) &=& \frac{2\pi}{3} \om \eta^4 e^4 \rho_0^4\left( \int_0^{E_{up}-2\hbar \om} dE f(E) \left( \ln\left|\frac{E_{up}-E-\hbar\om}{E+\hbar\om}\right|\right)^2 \right.\nonumber\\
&-& \left. \int_{2\hbar\om}^{E_{up}} dE f(E) \left( \ln\left|\frac{E_{up}-E+\hbar\om}{E-\hbar\om}\right|\right)^2 \right).
\label{6}
\end{eqnarray}
  \end{widetext}
And, with a  lengthy calculation, the fluctuation (equation (19) in supplementary information) can be simplified to 
\begin{eqnarray}
\la \dot{\hat{Q}}^4\ra &=&\frac{1}{6}\rho_0^4\eta^4 e^4 \om_{up}^7 \nonumber\\
&\times&\int_0^{E_{up}}dE
e^{-\beta E}(\frac{1}{3}\gamma^7 +
\frac{4}{3}\gamma^6+\frac{6}{3}\gamma^5+\frac{4}{3}\gamma^4+\frac{1}{3}\gamma^3)\nonumber\\.
\label{7}
\end{eqnarray}
Here $\gamma = \frac{E}{E_{up}}$.
Let us define
\begin{equation}
R_{Callen-Welton}(T)= \frac{\la \dot{\hat{Q}}^4\ra}{\int_0^\infty d\om Y_R^{(4)}(\om)},
\label{cwr}
\end{equation}
and call it Callen-Welton's ratio.  $R_{Callen-Welton} \propto T$ establishes the Nyquist's
relation. This is plotted in figure 2 as a dashed line (the solid line is for Nyquist's original
result). We notice that at high temperatures ($k_B T \gg \hbar\om$) we have the following relation
for our first example
\begin{equation}
\la \dot{\hat{Q}}^4\ra  \propto k_B T \int_0^\infty d\om Y_R^{(4)}(\om).
\label{nlnt}
\end{equation}
Next, we show that the above relation also holds good in our second system. The DOS model is the
same as before but the matrix elements are calculated by considering free electron gas in a 1-D
"box" of length $L$. Periodic boundary condition is applied to the plane wave state wavefunctions
$\phi_n(x) = \frac{1}{\sqrt{L}} e^{i k_n x}$ to mimic thermodynamic limit (later $L\rightarrow
\infty$ limit is taken). The matrix element $\la E_i|\hat{Q}|E_j\ra$ is equal to
$\frac{e\eta}{\sqrt{E_j}-\sqrt{E_i}}$ (see supplementary information).

The expression for $Y_R^{(4)}(\om)$ in this case is quite lengthy and is given in the supplementary information (equation (20)) but the fluctuations can be expressed by a simple formula
\begin{figure}[h!]
\includegraphics[height = 5cm, width = 7cm]{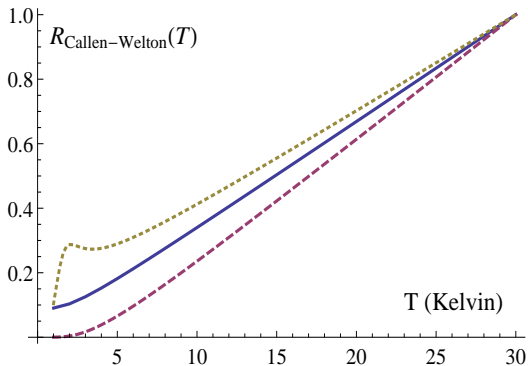}
\caption{Normalized Callen-Welton ratio $\frac{R_{Callen-Welton}(T)}{R_{Callen-Welton}(T=30~K)}$ as a function of temperature $T ~(Kelvin)$. Solid line is the original Nyquist's theorem. Dashed and dotted lines represents nonlinear Nyquist theorem. Notice the linearity of Callen-Welton's ratios for $T\gtrsim 5~K$.}
\label{f2}
\end{figure}
\begin{eqnarray}
\la\dot{\hat{Q}}^4\ra &=& \frac{8 e^4\rho_0^4\eta^4}{m^2\hbar^3}\int_0^{E_{up}} dE f(E) E^2
(E+E_{up}) \nonumber\\
&\times&\left(E_{up}^2 \log\left(\frac{E_{up}}{E_{up}-E}\right) + E^2 \log\left(\frac{E_{up}-E}{E}\right)\right).
\end{eqnarray}
Again, the Callen-Welton's ratio is plotted in figure 2 (dotted line) and we observe the validity of equation (\ref{nlnt}) in this case too\footnote{In figure 2 we plot normalized $R_{Callen-Welton}$ as we are not interested in the proportionality constants.}. This relation (equation (\ref{nlnt})) which is an exact analogue to the linear Nyquist theorem should be valid in general and can be dubbed as the {\it nonlinear Nyquist theorem}. Thus we conjecture that the equation (\ref{nlnt}) is a universal relation, an extension of the linear Nyquist relation to nonlinear regime. We further conjecture that
\begin{equation}
\la \dot{\hat{Q}}^{2 n}\ra  \propto k_B T \int_0^\infty d\om Y_R^{(n)}(\om).
\end{equation}
 $n=1$ is the Nyquist theorem and $n\geq 2$ are its nonlinear extensions!
Thus one can in principle compute nonlinear admittances from the character of {\it equilibrium
fluctuations}. Our conjectured relation(s) should be tested in other physical models, both
theoretically and experimentally. The conjecture is likely to be true in real physical systems as it
is valid in our considered examples especially in second physical system that has plane wave states.
Plane wave states is a reasonable approximation to electronic states in {\it real} metals. These
compact and explicit relations should have great practical applicability.

\vspace{2cm}

{\Huge Supplementary Information}

        \vspace{2cm}

All the technical details are collected here.

The total Hamiltonian is $H = H_0+V(t)\hat{Q}$. Where $\hat{Q} = \sum_i e \frac{\hat{x}_i}{L}$. Let eigen energies ($E_n$) and  eigenfunctions ($\phi_n$) of $H_0$ are known.

We want to compute an average power absorbed by the conductor from the battery. For this, let the wavefunction of the perturbed system is $\psi(t) =\sum_n a_n(t)\phi_n$ which obey the Schroedinger's equation $i \hbar \frac{\pr \psi(t)}{\pr t} = (H_0+ V(t)\hat{Q}) \psi(t)$. To the second order in perturbation theory, the expansion coefficient $a_n(t)$ of $\psi(t) =\sum_n a_n(t)\phi_n$  can be written as
\begin{widetext}
\begin{equation}
b_n(t) = (-i/\hbar)^2\sum_m \int_0^t dt^\p H^\p_{nm}(t^\p) \int_0^{t^\p} dt^{\p\p}
H^\p_{mi}(t^{\p\p}) e^{-i(\om_{mn}t^\p+\om_{in}t^{\p\p})} .
\end{equation}
\end{widetext}
With the transformation $a_n(t) = e^{-i E_n t/\hbar} b_{n}(t)$. Here $ H^\p_{nm}(t) = V_0 \sin(\om t) \hat{Q}$. The transition probability ($\mathscr{P}_{fi} =|\la f|\psi(t)\ra|^2$) takes the form:
\begin{widetext}
\begin{eqnarray}
\mathscr{P}_{fi} &=& (V_0/\hbar)^4\sum_{m,n} Q_{fn} Q_{ni} Q^\ast_{fm} Q^\ast_{mi} \int_0^t dt^\p \sin(\om t^\p) e^{-i \om_{nf} t^\p} \int_0^{t^\p} dt^{\p\p} \sin(\om t^{\p\p}) e^{- i \om_{in} t^{\p\p}} \nonumber\\
&\times&
\int_0^t dt^\p \sin(\om t^\p) e^{-i \om_{mf} t^\p} \int_0^{t^\p} dt^{\p\p} \sin(\om t^{\p\p}) e^{- i \om_{im} t^{\p\p}}.
\label{3}
\end{eqnarray}
 \end{widetext}
After performing the integrals in equation (\ref{3}) the transition probability can be written as
\begin{widetext}
\begin{eqnarray}
\mathscr{P}_{fi} &=& (V_0/2\hbar)^4\sum_{m,n} Q_{fn} Q_{ni} Q^\ast_{fm} Q^\ast_{mi} \nonumber\\
&\times& \left( \frac{e^{i(2\om+\om_{fi}) t}-1}{(\om-\om_{in})(2\om+\om_{fi})} - \frac{e^{i(\om-\om_{nf})t}-1}{(\om-\om_{in})(\om -\om_{nf})} - \frac{e^{-i\om_{if}t}-1}{(\om+\om_{in})\om_{if}} -\frac{e^{i(\om-\om_{nf})t}-1}{(\om+\om_{in})(\om-\om_{nf})} \right.\nonumber\\
&+& \left.\frac{e^{-i\om_{if} t}-1}{(\om-\om_{in})\om_{if}} - \frac{e^{-i(\om+\om_{nf})t}-1}{(\om-\om_{in})(\om +\om_{nf})} + \frac{e^{-i( 2\om+\om_{if})t}-1}{(\om+\om_{in})(2\om + \om_{if})} -\frac{e^{-i(\om+\om_{nf})t}-1}{(\om+\om_{in})(\om+\om_{nf})}  \right)\nonumber\\
&\times&\left( \frac{e^{-i(2\om+\om_{fi}) t}-1}{(\om-\om_{im})(2\om+\om_{fi})} - \frac{e^{-i(\om-\om_{mf})t}-1}{(\om-\om_{im})(\om -\om_{mf})} - \frac{e^{i\om_{if}t}-1}{(\om+\om_{im}\om_{if})} -\frac{e^{-i(\om-\om_{mf})t}-1}{(\om+\om_{im})(\om-\om_{mf})}   \right.\nonumber\\
&+& \left.  \frac{e^{i\om_{if} t}-1}{(\om-\om_{im})\om_{if}} - \frac{e^{i(\om+\om_{mf})t}-1}{(\om-\om_{im})(\om +\om_{mf})} + \frac{e^{i( 2\om+\om_{if})t}-1}{(\om+\om_{im})(2\om + \om_{if})} -\frac{e^{i(\om+\om_{mf})t}-1}{(\om+\om_{im})(\om+\om_{mf})}\right)\nonumber\\
\label{22}
\end{eqnarray}
  \end{widetext}
There are total $64$ terms, not all contribute in the long time limit: when $t$ is much greater than
 a characteristic time scale in the system and oscillation period of the applied field i.e.,
$t\gg\frac{1}{|\om_{if}|}$ and $t\gg\frac{1}{\om}$. Using the identity
\begin{equation}
lim_{t\rightarrow\infty} \left|\frac{e^{i x t}-1}{x}\right|^2 = 2\pi t \delta(x),
\end{equation}
Equation (\ref{22}) can be reduced to  equation (\ref{4}). Equation (\ref{5}) in the main text is
obtained as follows. First sums are converted  into integrals $\sum_n \longrightarrow \int dE
\rho(E)$.  Then integrals were performed over the final states using the properties of the Dirac
delta functions. This leads to total transition probability from initial state $|i\ra$ to {\it any
final state} per unit time i.e., $\mathscr{P}_i = \frac{1}{t}\sum_f \mathscr{P}_{if}$:
\begin{widetext}
\begin{eqnarray}
\mathscr{P}_i&=&(\pi V_0^4/8\hbar) \int_0^\infty dE_m \rho(E_m) \int_0^\infty dE_n
\rho(E_n)\nonumber\\
&\times& \left(\frac{\rho(E_i+2\hbar \om) Q_{E_i+2\hbar \om,E_n} Q_{E_n,E_i} Q^\ast_{E_i+ 2 \hbar \om,E_m} Q^\ast_{E_m,E_i} }{(\hbar \om +E_i -E_n)(\hbar\om +E_i-E_m)} \right. \nonumber \\ 
&+& \left. \frac{\rho(E_i-2\hbar \om) Q_{E_i-2\hbar \om,E_n} Q_{E_n,E_i} Q^\ast_{E_i-2 \hbar \om,E_m} Q^\ast_{E_m,E_i} }{(\hbar \om -E_i +E_n)(\hbar\om -E_i+E_m)}  \right)\nonumber\\
&+&(\pi V_0^4/2\hbar) \int_0^\infty dE_n \rho(E_n) \left(\frac{\hbar\om}{(\hbar\om)^2-(E_i-E_n)^2}\right)^2\nonumber\\
&\times& (\rho(E_n-\hbar\om)|Q_{E_n-\hbar\om,E_n}|^2|Q_{E_n,E_i}|^2+\rho(E_n+\hbar\om)|Q_{E_n+\hbar\om,E_n}|^2|Q_{E_n,E_i}|^2 )\nonumber\\.
\label{14}
\end{eqnarray}
\end{widetext}
The power absorbed by the conductor from battery (equation (\ref{5})) is calculated by subtracting the  loss from the gain (as done in Callen-Welton's formulation). Important point to notice in this nonlinear regime is that one has matrix elements of the form $\la E_i+2\hbar\om|\hat{Q}|E_n\ra\la E_n|\hat{Q}|E_i\ra$ which represents the absorption of {\it two quanta} from the battery in which initial state's energy changes from $E_i$ to $E_i+2 \hbar\om$ via some intermediate state of energy $E_n$ and the second matrix element $\la E_i-2\hbar\om|\hat{Q}|E_n\ra\la E_n|\hat{Q}|E_i\ra$ represents the loss of {\it two quanta}. The second term in the above equation (\ref{14}) in which only the intermediate states $|n\ra$ are changed does not contribute to the absorption of energy by the system from the battery because  the energy of the initial state $E_i$ in which the system was prepared at time $t=0$ remains unchanged. Thus net gain of energy by the system (final energy $-$ initial energy =$E_n\pm\hbar\om-E_i$) is zero when sum is performed on $n$ while in the first term we have: final energy$-$ initial energy $= E_i\pm2\hbar\om-E_i\ne 0$. Finally by performing an ensemble average over all possible initial states of energy $E_i$ (in which the system was prepared at time $t=0$)  with thermodynamical weighting factor $e^{-E_i/k_B T}$ leads to desired equation (\ref{5}). 

In linear regime the average power is given by $\la Power_\om\ra = \la\frac{V_0^2 \sin^2(\om t)}{R}\ra = \frac{V_0^2}{2R(\om)} = \frac{1}{2} V_0^2 Y(\om)$. In nonlinear regime it is $\frac{1}{2} V_0^2 Y(\om) + \frac{3}{8} V_0^4 Y_R^{(4)}(\om)+...$ with real part of the nonlinear admittance  given by $\la Power_\om\ra^{(4)} = \frac{3}{8} V_0^4 Y_R^{(4)}(\om)$ 
\begin{widetext}
\begin{eqnarray}
Y_R^{(4)}(\om) &=& \frac{2\pi}{3}\om \int_0^\infty dE_i \rho(E_i) f(E_i) \int_0^\infty dE_m
\rho(E_m) \int_0^\infty dE_n \rho(E_n)\nonumber\\
&\times& \left(\frac{\rho(E_i+2\hbar \om) Q_{E_i+2\hbar \om,E_n} Q_{E_n,E_i} Q^\ast_{E_i+ 2 \hbar \om,E_m} Q^\ast_{E_m,E_i} }{(\hbar \om +E_i -E_n)(\hbar\om +E_i-E_m)} \right. \nonumber \\ 
&-& \left. \frac{\rho(E_i-2\hbar \om) Q_{E_i-2\hbar \om,E_n} Q_{E_n,E_i} Q^\ast_{E_i-2 \hbar \om,E_m} Q^\ast_{E_m,E_i} }{(\hbar \om -E_i +E_n)(\hbar\om -E_i+E_m)}  \right).
\label{15}
\end{eqnarray}
\end{widetext}
This is an important and useful expression for $Y_R^{(4)}(\om)$ that can be used to calculate nonlinear admittance once matrix elements and DOS of a system are known!

The fluctuations are calculated according to Callen-Welton's formulation.  Thermal and quantum mechanical averages of fourth power of current operator are $\la \dot{\hat{Q}}^4\ra =\sum_n f(E_n) \la E_n| \dot{\hat{Q}}^4 |E_n\ra$. By inserting complete set of states one will have 

\begin{equation}
\la E_n|\dot{\hat{Q}}^4|E_n\ra = \sum_{m,p,q}\la E_n| \dot{\hat{Q}} |E_m \ra\la E_m| \dot{\hat{Q}} |E_p\ra\la E_p| \dot{\hat{Q}} |E_q \ra\la E_q| \dot{\hat{Q}} |E_n\ra.
\end{equation}
Using $\dot{\hat{Q}} = \frac{i}{\hbar}[H_0,\hat{Q}]$ and in parallel with Callen-Welton's formulation,  the quantum mechanical expectation value of the fluctuation takes the form
\begin{widetext}
\begin{eqnarray}
&&\la E_n |\dot{\hat{Q}}^4| E_n\ra  =\nonumber\\
&-&\int_0^\infty d\om \om \int_0^\infty d\om^\p (\om^\p -\om) \int_0^\infty d\om^{\p\p} (\om^{\p\p}-\om^\p) \rho(E_n-\hbar\om)\rho(E_n-\hbar\om^{\p})\rho(E_n-\hbar\om^{\p\p})\nonumber\\
&\times& Q_{E_n-\hbar\om^{\p},E_n+\hbar\om^{\p\p}} Q_{E_n+\hbar\om^{\p\p},E_n} Q_{E_n,E_n-\hbar\om} Q_{E_n-\hbar\om,E_n-\hbar\om^\p}\nonumber\\
&-&\int_0^\infty d\om \om \int_0^\infty d\om^\p (\om^\p -\om) \int_0^\infty d\om^{\p\p} (\om^{\p\p}+\om^\p) \rho(E_n-\hbar\om)\rho(E_n-\hbar\om^{\p})\rho(E_n+\hbar\om^{\p\p})\nonumber\\
&\times& Q_{E_n-\hbar\om^{\p},E_n-\hbar\om^{\p\p}} Q_{E_n-\hbar\om^{\p\p},E_n} Q_{E_n,E_n-\hbar\om} Q_{E_n-\hbar\om,E_n-\hbar\om^\p}\nonumber\\
&+&\int_0^\infty d\om \om \int_0^\infty d\om^\p (\om^\p +\om) \int_0^\infty d\om^{\p\p} (\om^{\p\p}+\om^\p) \rho(E_n-\hbar\om)\rho(E_n+\hbar\om^{\p})\rho(E_n-\hbar\om^{\p\p})\nonumber\\
&\times& Q_{E_n+\hbar\om^{\p},E_n+\hbar\om^{\p\p}} Q_{E_n+\hbar\om^{\p\p},E_n} Q_{E_n,E_n-\hbar\om} Q_{E_n-\hbar\om,E_n+\hbar\om^\p}\nonumber\\
&+&\int_0^\infty d\om \om \int_0^\infty d\om^\p (\om^\p +\om) \int_0^\infty d\om^{\p\p} (\om^{\p\p}-\om^\p) \rho(E_n-\hbar\om)\rho(E_n+\hbar\om^{\p})\rho(E_n+\hbar\om^{\p\p})\nonumber\\
&\times& Q_{E_n+\hbar\om^{\p},E_n-\hbar\om^{\p\p}} Q_{E_n-\hbar\om^{\p\p},E_n} Q_{E_n,E_n-\hbar\om} Q_{E_n-\hbar\om,E_n+\hbar\om^\p}\nonumber\\
&+&\int_0^\infty d\om \om \int_0^\infty d\om^\p (\om^\p +\om) \int_0^\infty d\om^{\p\p} (\om^{\p\p}-\om^\p) \rho(E_n+\hbar\om)\rho(E_n-\hbar\om^{\p})\rho(E_n-\hbar\om^{\p\p})\nonumber\\
&\times& Q_{E_n-\hbar\om^{\p},E_n+\hbar\om^{\p\p}} Q_{E_n+\hbar\om^{\p\p},E_n} Q_{E_n,E_n+\hbar\om} Q_{E_n+\hbar\om,E_n-\hbar\om^\p}\nonumber\\
&+&\int_0^\infty d\om \om \int_0^\infty d\om^\p (\om^\p +\om) \int_0^\infty d\om^{\p\p} (\om^{\p\p}+\om^\p) \rho(E_n+\hbar\om)\rho(E_n-\hbar\om^{\p})\rho(E_n+\hbar\om^{\p\p})\nonumber\\
&\times& Q_{E_n-\hbar\om^{\p},E_n-\hbar\om^{\p\p}} Q_{E_n-\hbar\om^{\p\p},E_n} Q_{E_n,E_n+\hbar\om} Q_{E_n+\hbar\om,E_n-\hbar\om^\p}\nonumber\\
&-&\int_0^\infty d\om \om \int_0^\infty d\om^\p (\om^\p -\om) \int_0^\infty d\om^{\p\p} (\om^{\p\p}+\om^\p) \rho(E_n+\hbar\om)\rho(E_n+\hbar\om^{\p})\rho(E_n-\hbar\om^{\p\p})\nonumber\\
&\times& Q_{E_n+\hbar\om^{\p},E_n+\hbar\om^{\p\p}} Q_{E_n+\hbar\om^{\p\p},E_n} Q_{E_n,E_n+\hbar\om} Q_{E_n+\hbar\om,E_n+\hbar\om^\p}\nonumber\\
&-&\int_0^\infty d\om \om \int_0^\infty d\om^\p (\om^\p -\om) \int_0^\infty d\om^{\p\p} (\om^{\p\p}-\om^\p) \rho(E_n+\hbar\om)\rho(E_n+\hbar\om^{\p})\rho(E_n+\hbar\om^{\p\p})\nonumber\\
&\times& Q_{E_n+\hbar\om^{\p},E_n-\hbar\om^{\p\p}} Q_{E_n-\hbar\om^{\p\p},E_n} Q_{E_n,E_n+\hbar\om} Q_{E_n+\hbar\om,E_n+\hbar\om^\p}
\end{eqnarray}
\end{widetext}
Due to this horrendous analytical complexity, it is not possible to establish a direct relation between $Y_R^{(4)}(\om)$ and $\la \dot{\hat{Q}}\ra$. Thus the above expressions are simplified in simple specific models. In our first model with constant DOS and constant matrix elements the expressions for nonlinear admittance and fluctuation are simpler and are given in equations (\ref{6}) and (\ref{7}) respectively (in the main text). 

For our second model the nonlinear admittance takes the form:
\begin{widetext}
\begin{eqnarray}
&&Y_R^{(4)}=\frac{2\pi}{3} \om \eta^4 e^4  \rho_0^4\nonumber\\
&\times&\int_0^{E_{up-2\hbar\om}} dE f(E) \left(\frac{\ln[(f_c(E_{up})-f_a(E))/f_a(E)]}{2 (f_a(E)-f_b(E))(f_a(E)-f_c(E))}+\frac{\ln[(f_c(E_{up})+f_a(E))/f_a(E)]}{ 2 (f_a(E)+f_b(E))(f_a(E)+f_c(E)) }\right.\nonumber\\
&-& \left. \frac{f_b(E)\ln[(f_c(E_{up})-f_b(E))/f_b(E)]}{(f_a^2(E)-f_b^2(E))(f_b(E)-f_c(E))}-\frac{f_c(E)\ln[(f_c(E_{up})-f_c(E))/f_c(E)]}{(f_a^2(E)-f_c^2(E))(f_c(E)+f_b(E)) }\right)^2\nonumber\\
&-& \frac{2\pi}{3} \om \eta^4 e^4  \rho_0^4\int_{2\hbar \om}^{E_{up}} dE f(E) \left(....f_a,~f_b....replaced~with....\bar{f}_a,~\bar{f}_b....\right)^2.
\end{eqnarray}
 \end{widetext}
Here $f_a(E) = \sqrt{E+\hbar\om},~f_b(E) = \sqrt{E+2\hbar\om},~f_c(E) = \sqrt{E}$ and $\bar{f}_a(E) = \sqrt{E-\hbar\om},~\bar{f}_b(E) = \sqrt{E-2\hbar\om}$. In low frequency regime when $\hbar\om\ll E_{up}$ and $\hbar\om\ll k_B T$ the above expression takes a simpler form:
\begin{equation}
Y_R^{(4)} \sim \frac{(\hbar \om)^2}{k_B T}\log(\frac{E_{up}}{2\hbar\om})\int_{2\hbar\om}^{E_{up}} dx \frac{e^{-x/k_B T}}{x^2}
\end{equation}
This is in sharp contrast to the Drude form $\sigma(\om)\sim 1/(1+\om^2 \tau^2)$ of linear admittance ($\tau$ is the Drude scattering rate). In the low frequency limit  ($\om^2\tau^2\ll 1$) Drude conductivity (which is proportional to linear admittance) is approximately constant and then decrease with frequency whereas the nonlinear admittance increases with frequency in the low frequency limit (ref to above equation).  The matrix elements in our second model are computed as follows:
  \begin{widetext}
\begin{equation}
\la E_i|\hat{Q}| E_j \ra = \frac{e}{L} \sum_{l=1}^N \la E_i|\hat{x}_l| E_j \ra = \sum_{l=1}^N\sum_{j,j^\p} \int dx_j \int dy_{j^\p} \la E_n | x_j\ra\la x_j|\hat{x}_l|y_{j^\p}\ra\la y_{j^\p}| E_m\ra.
\end{equation}
 \end{widetext}
Here $N$ is the number of particles in the system and $\la E_n|x_j\ra = \frac{1}{\sqrt{L}} e^{-i k_n x_j}$ are plane wave states. Using the properties of delta-functions resulting from  $\la x_j|\hat{x}_i|y_{j^\p}\ra = x_i \delta_{i,j^\p} \delta_{j,j^\p}\delta(x_i-y_{j^\p}) \delta(x_j-y_{j^\p})$ , the sums and integrals can be easily simplified. Further using the periodic boundary conditions $k_f-k_i=k_{fi} = 2 n (\pi/L)$ (to mimic a thermodynamic system), the matrix elements takes the form $ e \eta/(\sqrt{E_j} - \sqrt{E_i})$, where $\eta$ is a constant, and $E_i =\frac{\hbar^2 k_i^2}{2 m}$ for plane wave state.

\end{document}